\documentclass[11pt,a4paper]{article}
\pdfoutput=1
\usepackage{jcappub}
\graphicspath{{images/}}

\usepackage{comment}
\usepackage{amsmath,bm,accents}
\usepackage{bm,accents}
\usepackage{booktabs}
\usepackage{multirow}
\usepackage{mathtools}
\makeatletter
\newcommand\accbm[1]{\use@mathgroup{\M@OMS}{7}{\bm{#1}}}
\makeatother
\usepackage{physics}
\usepackage{cancel}
\usepackage{stackengine}
\usepackage{ulem}
\usepackage[textwidth=20mm]{todonotes}

\title{\boldmath Impact of adiabatic temperature fluctuations on the power spectrum of axion density perturbations}
\author[1]{Ahmed Ayad\note{Corresponding author.}}
\author{and Dominik J.~Schwarz}
\affiliation{Fakult\"at f\"ur Physik, Universit\"at Bielefeld,\\Postfach 100131, 33501 Bielefeld, Germany}

\emailAdd{ayad@physik.uni-bielefeld.de}
\emailAdd{dschwarz@physik.uni-bielefeld.de}
\abstract{Axions and axion-like particles (ALPs) have gained substantial attention as potential candidates for cold dark matter. The ALP field can exhibit fluctuations stemming from initial conditions. These initial field fluctuations hold the potential to give rise to gravitationally bound configurations known as axion miniclusters (AMC). While this proposition is widely accepted in the post-inflationary Peccei-Quinn symmetry-breaking scenario, uncertainties persist regarding the pre-inflationary scenario, where the effects of the initial field fluctuations may be suppressed by inflation. In this study, we investigate the influence of adiabatic temperature fluctuations of the primordial plasma on the evolution of axion density perturbations and their power spectrum, aiming to explore the possibility of AMC formation in the pre-inflationary scenario. Our analysis reveals that the impact of adiabatic temperature fluctuations becomes significant when $f_{\rm a} / H_{\rm inf} \gtrsim 1.25 \times 10^4$ and surpasses that of quantum fluctuations by up to five orders of magnitude on large scales. This emphasizes the possibility of also forming AMC within the pre-inflationary scenario. Consequently, a detection of AMC would not reliably differentiate between the pre- and post-inflationary origin of axions.}
\arxivnumber{2503.05532}
\begin{document}
\maketitle
\flushbottom

\section{Introduction} \label{sec:1}

The existence of pseudoscalar fields of axions~\cite{Weinberg:1977ma, Wilczek:1977pj} or axion-like particles (ALPs)~\cite{Anselm:1981aw, Kim:1986ax, Jaeckel:2010ni} has gained significant attention in both particle physics and cosmology. They have emerged as potential candidates for resolving a variety of fundamental problems, including the 
nature of cold dark matter (CDM)~\cite{Abbott:1982af, Preskill:1982cy, Dine:1982ah, Arias:2012az}. Extensive endeavors encompassing theory development, observational studies, and experimental investigations have been devoted to the pursuit of axions and ALPs~\cite{Marsh:2015xka, Sikivie:2020zpn, Raffelt:1990yz, Raffelt:2006cw, Sigl:2017wfx}.

QCD axions are Nambu-Goldstone bosons resulting from the spontaneous breaking of the Peccei-Quinn (PQ) symmetry~\cite{Peccei:1977hh, Peccei:1977ur} that provide a solution to the strong charge-parity (CP) problem~\cite{Cheng:1987gp}. The strong CP problem arises from the absence of observable CP symmetry violation in the strong interactions, as evidenced by the remarkably small value of the electric dipole moment of the neutron and the inferred upper limit of the vacuum angle\footnote{The structure of the vacuum state can be parameterized by the total vacuum angle $\bar{\theta}$, which receives contributions from both strong and weak interactions~\cite{Coleman:1985rnk, ParticleDataGroup:2014cgo}.}
$\theta \lesssim 5 \times 10^{-11}$~\cite{Pospelov:2005pr, Abel:2020pzs}. For comprehensive reviews on this topic, refer to Refs.~\cite{Sikivie:2006ni, Chadha-Day:2021szb}. On the other hand, ALPs are particles similar to QCD axions and arise in various theoretical models including string~\cite{Arvanitaki:2009fg, Cicoli:2012sz, Anselm:1981aw}, supersymmetry~\cite{Nilles:1981py}, and grand unified theories~\cite{Wise:1981ry}. They are proposed to address various issues in these models, primarily those not directly associated with the strong CP problem~\cite{Jaeckel:2010ni, Arias:2012az}. For detailed reviews, see~\cite{Irastorza:2018dyq, ParticleDataGroup:2022pth, Caputo:2024oqc}. The properties of axions and ALPs depend on an energy scale $f_{\rm a}$ at which the associated symmetry is spontaneously broken. The QCD axion-gluon coupling, required to solve the strong CP problem, directly links the axion mass $m_{\rm a}$ to the energy scale $f_{\rm a}$, but this relationship may not necessarily apply to generic ALPs.

In principle, ALPs, including QCD axions, can be produced through both thermal and non-thermal processes~\cite{Sikivie:1982qv}. Various mechanisms contribute to the production of a thermal population of ALPs~\cite{Turner:1986tb}. However, their role as cold dark matter is substantially restricted within the context of CDM models across most of the parameter space~\cite{Blumenthal:1984bp, Raffelt:2006rj, Sikivie:2006ni}. In the early Universe, their main production mechanisms are non-thermal and involve the misalignment mechanism~\cite{Preskill:1982cy, Dine:1982ah, Abbott:1982af}, and under specific conditions, the decay of topological defects such as axion strings and domain walls~\cite{Davis:1986xc, Lyth:1991bb}. The contribution of each mechanism to the relic density of axions hinges on whether the corresponding symmetry-breaking occurred before or after the inflationary epoch~\cite{Starobinsky:1980te, Kazanas:1980tx, Sato:1980yn, Guth:1980zm, Linde:1981mu, Albrecht:1982wi}. 
The misalignment mechanism arises from the initial misalignment of the axion field, while topological defects can form during phase transitions and subsequently decay, releasing more axions. 

In the pre-inflationary scenario, where the symmetry breaking scale $f_{\rm a}$ is larger than or comparable to the energy scale of inflation, the PQ symmetry breaks before the end of inflation. This implies that the axion field exists during the early accelerated expansion of the Universe. Due to the inflationary process, the observable Universe is contained within a causally connected patch, resulting in a single random initial value $\theta_{\rm i}$ for the axion misalignment field. As a consequence, the axion field becomes homogeneous over vast distances, including the entire observable Universe. Topological defects are expected to be inflated away and do not contribute significantly to the axion energy density. Therefore, in the pre-inflationary scenario, the dominant process that contributes to the energy density of axions is the misalignment mechanism~\cite{Preskill:1982cy, Abbott:1982af, Dine:1982ah}.

In the post-inflationary scenario, where the symmetry-breaking scale $f_{\rm a}$ is smaller than the energy scale of inflation, the two degrees of freedom of the PQ field reduce to the single degree of freedom of the axion field only after the inflationary period ends. In this case, the axion field also takes random initial values in causally disconnected regions. However, because the PQ symmetry breaking occurs after the end of inflation, these field fluctuations persist and serve as the source of initial perturbations in the axion field~\cite{Axenides:1983hj, Steinhardt:1983ia, Linde:1985yf, Seckel:1985tj, Lyth:1989pb, Turner:1990uz}. This leads to the formation of spatial fluctuations in the axion field across the Universe and the formation of topological defects. The energy density of axions in this scenario is primarily thought to be contributed by the misalignment mechanism, where the axion field typically settles into its minimum potential. Additional contributions arise from the decay of topological defects that were formed during the PQ symmetry breaking. In the post-inflationary case, the energy density of axions does not depend solely on the initial value of the axion field but rather on its overall dynamics.

The time evolution of initial field fluctuations in the post-inflationary scenario can give rise to significant fluctuations in the energy density distribution of the axion field. These density fluctuations have the potential to lead to the formation of gravitationally bound overdense configurations of axion miniclusters (AMC)~\cite{Hogan:1988mp}. Recent studies have provided detailed calculations on the power spectrum of these fluctuations as well as predictions of AMC properties~\cite{DiLuzio:2020wdo, Feix:2020txt, Feix:2019lpo, Enander:2017ogx, Tinyakov:2015cgg, Davidson:2016uok, Bai:2016wpg, OHare:2021zrq, Ellis:2020gtq, Eggemeier:2019khm}. Consequently, the presence of AMC has been proposed as a potential signature of the post-inflationary scenario, with potential detection methods including microlensing and femtolensing surveys and pulsar-timing arrays, which could serve as smoking gun evidence~\cite{Kolb:1995bu, Fairbairn:2017dmf, Fairbairn:2017sil, Dai:2019lud, Xiao:2024qay}.

Recent studies have revisited the possibility of AMC formation in the pre-inflationary scenario. Ref.~\cite{Arvanitaki:2019rax} demonstrated that attractive self-interactions in scalar field models, driven by parametric resonance during the radiation era, can lead to the formation of compact structures such as halos, solitons, and oscillons, significantly influencing density perturbations across a wide range of axion masses. In contrast, Ref.~\cite{Fukunaga:2020mvq} found that QCD axions do not form clumps even under fine-tuned initial conditions, while ALPs with more general potential forms can form clumps through tachyonic or resonance instabilities, particularly in the case of a multiple cosine potential.  However, these investigations initially overlooked the impact of adiabatic temperature fluctuations in QCD matter seeded by cosmological inflation \cite{Mukhanov:1981xt, Hawking:1982cz, Guth:1982ec, Bardeen:1983qw}, which could significantly alter the spatial distribution of axions at sub-galactic scales. Ref.~\cite{Kitajima:2021inh} highlighted the importance of direct interactions between axions and QCD matter during the QCD phase transition. It identified two mechanisms that can lead to AMC formation in the pre-inflationary scenario: enhanced primordial curvature perturbations at the QCD horizon scale and fine-tuned initial misalignment near the potential's hilltop. Additionally, Ref.~\cite{Sikivie:2021trt} explored how adiabatic temperature fluctuations seeded by inflation influence axion dynamics during the QCD phase transition, particularly focusing on their effects on the axion momentum distribution. These findings suggest significant implications for the spatial and kinematic properties of axions, especially for masses exceeding a few $\mu{\rm eV}$. Together, these studies emphasize the critical role of adiabatic temperature fluctuations in shaping the spatial distribution of axions at sub-galactic scales, with profound implications for the formation of primordial dense axion configurations.

In this study, our focus is to analyze the inhomogeneous dynamics of the axion field during the QCD epoch of the early Universe in the pre-inflationary scenario. Specifically we investigate the power spectrum of field fluctuations induced by adiabatic temperature fluctuations of the QCD matter. We examine how these adiabatic temperature fluctuations can influence the evolution of axion density perturbations. Our analysis challenges the previous belief that the presence of AMC alone can reliably differentiate between a pre- and post-inflationary origin of axions. 

This work is organized as follows. In Section~\ref{sec:2}, we explore the fundamental properties of the QCD axion and ALP field and their intersection with the thermal dynamics of the early Universe. Section~\ref{sec:3} provides a comprehensive analysis of the equation of motion governing the axion field, detailing the dynamics and perturbations. In Section~\ref{sec:4}, we revise the perturbed equation of motion, offering a detailed analysis of this key component. Subsequently, Section~\ref{sec:5} explores the energy density distribution and power spectrum of the axion field. In Section~\ref{sec:6}, we analyze our findings and discuss their implications in detail. Finally, in Section~\ref{sec:7}, we present our concluding remarks, summarizing the key outcomes and identifying potential avenues for future research.

\section{Axion field dynamics} \label{sec:2}

In this section, we discuss the relic abundance of axion CDM and explore its initial conditions. Furthermore, we construct a simplified model for the time and temperature dependence of axion mass.

\subsection{Axion field evolution in the thermal Universe} \label{sec:2.1}

The dynamics of the axion field are primarily governed by its potential, which is intricately linked to its mass, a temperature-dependent quantity. These dynamics, including the axion field oscillations, are particularly relevant in the context of its potential role as dark matter during the radiation-dominated epoch.

In the early Universe, when temperatures were high and the Universe was radiation-dominated, its expansion was governed by the Hubble rate~\cite{Kolb:1990vq}
\begin{equation} \label{eq:2.1}
H \equiv  \frac{\dot{R}}{R} = \sqrt{\frac{8 \pi G_{\rm N} \rho}{3}} \simeq 1.66 \sqrt{g_{\rm \ast}(T)} \frac{T^2}{M_{\rm Pl}} \,.
\end{equation}
Here, $R$ represents the scale factor of the Universe, $\dot{R}$ denotes the derivative of the scale factor with respect to cosmic time, $G_{\rm N}$ is the Newtonian gravitational constant, $\rho$ denotes the energy density of the Universe, $g_{\rm \ast}(T)$ reflects the number of relativistic degrees of freedom at temperature $T$, and $M_{\rm Pl}$ is the Planck mass. The energy density and entropy density of relativistic particles during the radiation-dominated era are given by 
\begin{equation} \label{eq:2.2}
\rho(T) = \frac{\pi^2}{30} g_{\rm \ast}(T) T^4 \,, \qquad  s(T) =\frac{2 \pi^2}{45} g_{\rm \ast s}(T) T^3 \,.
\end{equation} 
Here, $g_{\rm \ast s}(T)$ represents the effective number of relativistic degrees of freedom contributing to the entropy density. $g_{\rm \ast}(T)$ and $g_{\rm \ast s}(T)$ are generally similar but can differ due to how particles with different masses and statistics contribute to energy and entropy. Notably, around the QCD phase transition, where quarks and gluons form hadrons, $g_{\rm \ast}(T)$ and $g_{\rm \ast s}(T)$ can show noticeable differences~\cite{Kolb:1990vq}.

The conservation of entropy implies
\begin{equation}
\frac{{\rm d}s}{{\rm d}T} \frac{{\rm d}T}{{\rm d}t} = - 3 H s \,. 
\label{adiabatic_expansion}
\end{equation} 
Using Eqs.~\eqref{eq:2.1}, \eqref{eq:2.2}, and \eqref{adiabatic_expansion} with ${\rm d}\rho = T {\rm d}s$ and speed of sound $c_{\rm s} = [s/(T {\rm d} s/{\rm d} T)]^{1/2}$, for adiabatic processes, the temperature dependence of time in the early Universe can be determined as
\begin{equation} \label{eq:2.3}
\dfrac{{\rm d}t}{{\rm d}T} = - \frac{1}{3 c_{\rm s}^2 H T} = - M_{\rm Pl} \sqrt{\frac{45}{64 \pi^3}} \frac{1}{T^3 g_{\rm \ast s}(T) \sqrt{g_{\rm \ast}(T)}} \left( T \dfrac{{\rm d}g_{\rm \ast}(T)}{{\rm d}T} + 4 g_{\rm \ast}(T) \right) \,.
\end{equation}
This equation provides crucial insights into the thermal history and how the temperature evolves over time in the early Universe. For temperatures much greater than $250 \, {\rm GeV}$, above the electroweak symmetry breaking scale, $g_{\rm \ast}(T)$ is approximately $106.75$, accounting for all degrees of freedom of the standard model. However, for late times and low temperatures below $1 \, {\rm MeV}$, following primordial Big Bang Nucleosynthesis, $g_{\rm \ast}(T)$ equals $3.36$. In the intermediate phase between these two periods, particularly around the QCD phase transition scale, $g_{\rm \ast}(T)$ can be computed as a function of temperature. Further details can be found in Refs.~\cite{ParticleDataGroup:2014cgo, Kolb:1990vq, Laine:2016hma}.

\begin{figure}[t!]
\centering
\includegraphics[width=0.65\textwidth]{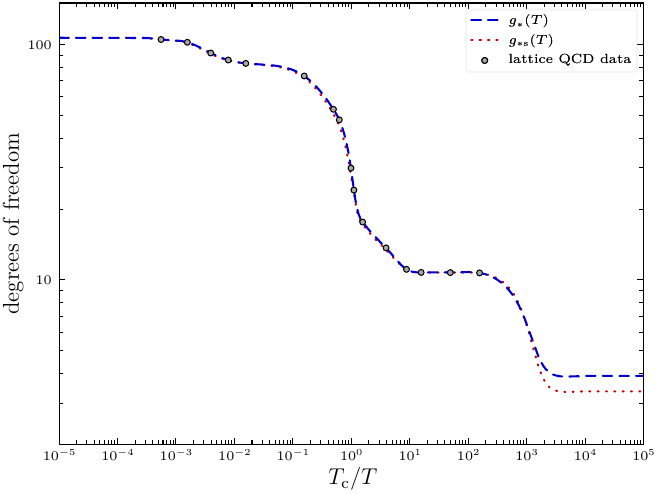}
\caption{The effective degrees of freedom for the energy density $g_{\rm \ast}(T)$ and entropy density $g_{\rm \ast s}(T)$ as functions of temperature $T$, calculated using lattice QCD data from Ref.~\cite{Borsanyi:2016ksw}.}
\label{fig:dof}
\end{figure}

Figure~\ref{fig:dof} illustrates the number of effective relativistic degrees of freedom for the energy density $g_{\rm \ast}(T)$ and entropy density $g_{\rm \ast s}(T)$ as functions of temperature $T$, calculated using lattice QCD data from Ref.~\cite{Borsanyi:2016ksw}. Understanding the behavior of $g_{\rm \ast}(T)$ during the QCD phase transition is crucial for accurately modeling the evolution of the axion field. The QCD phase transition, occurring at a pseudo-critical temperature $T_{\rm c} = 0.1565(15) \, {\rm GeV}$~\cite{HotQCD:2018pds}, results in the confinement of quarks and gluons within hadrons and significantly reduces their degrees of freedom.

We note that the conversion from cosmic time to temperature in Eq.~\eqref{eq:2.3} incorporates the temperature dependence of the effective number of relativistic degrees of freedom $g_{\rm \ast}(T)$ and the entropy degrees of freedom $g_{\rm \ast s}(T)$. This dependence becomes especially relevant around the QCD phase transition, where both $g_{\rm \ast}(T)$ and $g_{\rm \ast s}(T)$ vary significantly. Since the axion mass is a temperature-dependent quantity arising from QCD effects, the evolution of the axion field is indirectly influenced through both the expansion rate and the damping term in its equation of motion. These effects are particularly important when evaluating axion density fluctuations sourced by temperature inhomogeneities near the QCD epoch. Moreover, they play a role in determining the onset of oscillations and the final relic abundance generated via the misalignment mechanism~\cite{DiLuzio:2020wdo}. Throughout this work, $g_{\rm \ast}(T)$ and $g_{\rm \ast s}(T)$ are consistently included in our numerical treatment via the time-temperature relation. For a discussion of their impact on the evolution of axion fluctuations, see Section~\ref{sec:3}.

\subsection{Axion mass and relic density} \label{sec:2.2}

The mass evolution of the axion is pivotal in understanding its behavior as it transitions from a massless state to a massive one. Initially, at very high temperatures ($T \gg T_{\rm c}$) -- but in the PQ-symmetry broken phase, axions are essentially massless. As the Universe cools, particularly around the time of the QCD phase transition ($T \sim T_{\rm c}$), approximately $10 \, {\rm \mu s}$ after the Big Bang, non-perturbative QCD effects, such as instantons, generate an axion potential. This potential causes the axion field to oscillate around its minimum, resulting in a non-zero axion mass~\cite{Vafa:1984xg}. The temperature-dependent mass of the axion field $m_{\rm a}$ can be approximated by leveraging the QCD topological susceptibility, which is described as follows~\cite{Borsanyi:2016ksw}
\begin{equation} \label{eq:2.4}
m_{\rm a}(T) = \frac{\sqrt{\chi(T)}}{f_{\rm a}}, \quad \chi(T)= \frac{\chi_{\rm 0}}{1+(T/T_{\rm c})^b} \,,
\end{equation}
where $\chi_{\rm 0} = (7.55(5) \times 10^{-2} \, {\rm GeV})^4$~\cite{GrillidiCortona:2015jxo} represents the topological susceptibility at zero temperature, $b= 8.16$~\cite{Borsanyi:2016ksw} characterizes the power-law behavior as temperature varies, and $T_{\rm c}$ signifies the critical temperature for the QCD phase transition.

As the temperature decreases further ($T \ll T_{\rm c}$), axions reach their respective zero-temperature masses $m_{\rm 0}$. For QCD axions, the zero-temperature mass is approximately given by~\cite{Weinberg:1977ma}
\begin{equation} \label{eq:2.5}
m_{\rm 0} \simeq \frac{m_{\rm \pi} f_{\rm \pi}}{f_{\rm a}} \frac{\sqrt{m_{\rm u}m_{\rm d}}}{m_{\rm u} + m_{\rm d}} \simeq 5.70(7) \times 10^{-10} \left( \frac{10^{16} \, {\rm GeV}}{f_{\rm a}} \right) \, {\rm eV} \,,
\end{equation}
where $m_{\rm \pi}$ and $f_{\rm \pi}$ represent the pion mass and decay constant, respectively, and $m_{\rm u}$ and $m_{\rm d}$ are the up and down quark masses. This transition behavior in axion mass plays a central role in the study of axion dynamics.

Presently, astrophysical constraints provide stringent boundaries on the range of the energy scale $f_{\rm a}$, with established lower and upper limits. The lower limit, based on the stellar cooling argument for supernova 1987A, places $f_{\rm a}$ above $4 \times 10^{8} \, {\rm GeV}$~\cite{Mayle:1987as}. The upper limit, derived from the absence of observable effects of light axion condensates on the dynamics of smaller stellar mass black holes, is around $3 \times 10^{17} \, {\rm GeV}$~\cite{Arvanitaki:2009fg, Arvanitaki:2010sy, Arvanitaki:2014wva}. The relic abundance of axions in the current epoch can then be determined by the initial misalignment angle $\theta_{\rm i}$ (see Section~\ref{sec:3} for its definition) and the energy scale $f_{\rm a}$, as described by~\cite{Visinelli:2009zm}
\begin{equation} \label{eq:2.6}
\Omega_{\rm a} h^2 \simeq
\begin{dcases}
1.16 \times 10^4 \, \theta_{\rm i}^2 \left( \frac{f_{\rm a}}{10^{16} \, {\rm GeV}} \right)^{7/6} \,, \:\: \qquad f_{\rm a} \lesssim \hat{f}_{\rm a} \,, \\
0.54 \times 10^4\, \theta_{\rm i}^2 \left( \frac{f_{\rm a}}{10^{16} \, {\rm GeV}} \right)^{2/3} \,, \qquad f_{\rm a} \gtrsim \hat{f}_{\rm a} \,,
\end{dcases}
\end{equation}
where $\Omega_{\rm a}$ is the fractional contribution of axions to the total energy density, $h$ is the value of the current Hubble constant in units of $100 \, {\rm km} \, {\rm s}^{-1} \, {\rm Mpc}^{-1}$, and $\hat{f}_{\rm a} = 9.91 \times 10^{16} \, {\rm GeV}$. In the pre-inflationary scenario, $f_{\rm a} \sim 10^{12} \, {\rm GeV}$ is a popular choice such that $\theta_{\rm i}$ is of the order of unity to account for the total observed dark matter content in the Universe as shown in Eq.~\eqref{eq:2.6}. Another intriguing option is to set $f_{\rm a}$ at the grand unification scale about $\sim 10^{16} \, {\rm GeV}$, which requires $\theta_{\rm i}$ to be of the order of $\approx 3.2 \times 10^{-3}$ to explain the overall dark matter abundance in the Universe~\cite{Abbott:1982af}. In the post-inflationary scenario, the initial misalignment angle $\theta_{\rm i}$ typically converges to around $\theta_{\rm i} \approx 2.155$\footnote{For a quadratic potential, the average initial condition is $\theta_{\rm i} \approx \pi/\sqrt{3} \simeq 1.81$. However, for the QCD axion, anharmonic corrections modify this value, leading to $\theta_{\rm i} \approx 2.155$~\cite{GrillidiCortona:2015jxo}.}, reflecting the average taken over all values of $\theta_{\rm i} \in [-\pi, \pi]$ within a Hubble patch~\cite{Borsanyi:2016ksw}. This average serves as a pivotal convergence point for estimating the axion's contribution to CDM within the post-inflationary regime.

\begin{figure}[t!]
\centering
\includegraphics[width=0.65\textwidth]{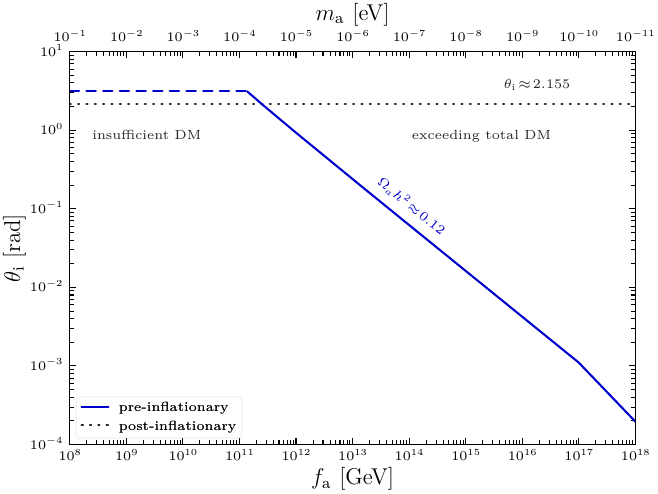}
\caption{The plot showcases the relation between the energy scale $f_{\rm a}$ and the initial misalignment angle $\theta_{\rm i}$, where the energy density from the misalignment mechanism matches the dark matter density, obtained by solving Eq.~\eqref{eq:2.6}.}
\label{fig:initial-angle}
\end{figure}

Figure~\ref{fig:initial-angle} depicts the relation between the energy scale $f_{\rm a}$ and the initial misalignment angle $\theta_{\rm i}$. The black dotted line represents the post-inflationary case, while the blue solid line represents the pre-inflationary case, accounting for the total dark matter made of axions produced by the misalignment mechanism. Regions above the blue solid line indicate an overproduction of dark matter, while regions below it correspond to the production of insufficient axion dark matter relic. This representation offers valuable insights into how varying values of $f_{\rm a}$ and $\theta_{\rm i}$ influence the axion's contribution to dark matter. In the following we focus on the pre-inflationary scenario.

\subsection{Simplified axion mass model} \label{Sec:2.3}

To model the mass acquisition of axions, we can make a simple approximation by assuming that axions suddenly acquire their masses at a specific moment $t_{\rm a}$ of temperature $T_{\rm a}$, which is justified by the rather sharp onset of the axion mass, see Eq.~(\ref{eq:2.4}). Above a characteristic temperature $T_{\rm a}$, the axion mass remains zero, while below this temperature, it takes on the value of the zero-temperature mass $m_{\rm 0}$. Therefore, we can express the squared axion mass as follows
\begin{equation} \label{eq:2.7}
m_{\rm a}^2(T) = m^2_{\rm 0} \, \Theta (T_{\rm a} - T) \,,
\end{equation}
where $\Theta(x)$ represents the Heaviside step function. This approximation simplifies the treatment of the temperature dependence of the axion mass and allows us to analyze its effects more conveniently.

\begin{figure}[t!]
\centering
\includegraphics[width=0.4963\textwidth]{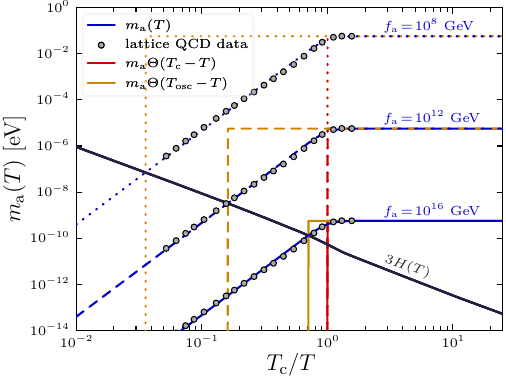}
\includegraphics[width=0.4963\textwidth]{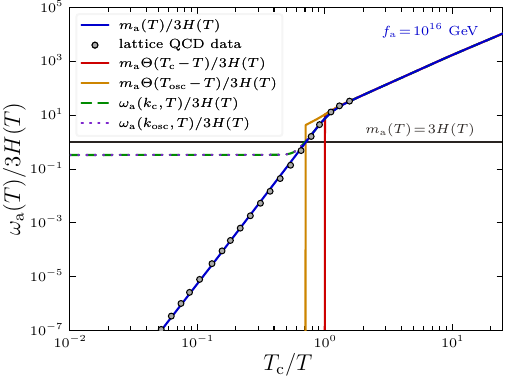}
\caption{The left panel plot depicts the behavior of the axion mass $m_{\rm a}(T)$ across a range of temperatures. The blue curves represent the temperature-dependent axion mass, incorporating insights from QCD topological susceptibility calculations using lattice QCD data from Ref.~\cite{Borsanyi:2016ksw}. These curves are drawn for different PQ breaking scales: $f_{\rm a} = 10^{8} \, {\rm GeV}$, $f_{\rm a} = 10^{12} \, {\rm GeV}$, and $f_{\rm a} = 10^{16} \, {\rm GeV}$. The red curves provide a simplified perspective, assuming an instantaneous acquisition of mass by axions at the critical temperature $T_{\rm c}$. Moreover, the orange curves introduce a scenario where axions acquire mass abruptly at $T_{\rm osc}$, coinciding with the onset of axion field oscillations. The right panel plot illustrates the ratio of $\omega_{\rm a}(T)/3H(T)$ as a function of temperature for the regime of $f_{\rm a} = 10^{16} \, {\rm GeV}$ with lattice QCD data from Ref.~\cite{Borsanyi:2016ksw}. The black line marks the point where the axion mass $m_{\rm a}$ becomes significant, defined as $m_{\rm a} = 3H$. This threshold serves as a crucial indicator of the significance of the axion mass concerning the expansion rate of the Universe. Additionally, this panel includes $\omega_{\rm a}(k,T) / 3H(T)$ for the two modes: $k_{\rm c} = R(T_{\rm c}) \, H(T_{\rm c})$ and $k_{\rm osc} = R(T_{\rm osc}) \, H(T_{\rm osc})$. These serve as reference points when comparing these modes to the axion mass.}
\label{fig:mass}
\end{figure}

In the left panel of Figure~\ref{fig:mass}, we illustrate the behavior of the axion mass $m_{\rm a}(T)$ over a range of temperatures. This analysis is conducted with PQ breaking scales set at $f_{\rm a} = 10^{8} \, {\rm GeV}$, $f_{\rm a} = 10^{12} \, {\rm GeV}$, and $f_{\rm a} = 10^{16} \, {\rm GeV}$. The blue curves represent the temperature-dependent axion mass, derived from QCD topological susceptibility calculations based on lattice simulation data found in Ref.~\cite{Borsanyi:2016ksw}. These curves depict the gradual increase in axion mass as the temperature decreases, signifying the transition from a massless state to a massive one, as described by Eq.~\eqref{eq:2.4}. In contrast, the red curves adopt a simpler approach, assuming that axions instantaneously acquire mass at the critical temperature $T_{\rm c}$. This simplification offers a useful estimation of axion mass behavior without delving into the intricate details of the QCD phase transition. However, the axion field starts its oscillations when $m_{\rm a} \sim 3 H$, which happens when the Universe temperature drops to $T_{\rm osc}$. We therefore also show a scenario where axions swiftly gain mass at $T_{\rm osc}$, coinciding with the onset of axion field oscillations (orange curves). This streamlined approach provides valuable insights into the evolution and behavior of axion field perturbations, offering a practical estimation of axion mass behavior without the need to consider the intricacies of the QCD phase transition. A comparison between the simplified model and the full temperature dependence of the axion mass reveals reasonable agreement, justifying the use of this simplified assumption in revising the perturbed equation of motion to account for adiabatic temperature fluctuations in Section~\ref{sec:4}.

In the right panel of Figure~\ref{fig:mass}, we consider spatial axion field fluctuations with wavenumber $k$ and plot the ratio between $\omega_{\rm a}(k,T) = [k^2/R^2(T) + m_{\rm a}^2(T)]^{1/2}$ and $3H$ across various temperatures for the regime with $f_{\rm a} = 10^{16} \, {\rm GeV}$. The blue line captures $m_{\rm a}(T)/3H(T)$ where $m_{\rm a}(T)=\omega_{\rm a}(0,T)$. This graph identifies a crucial threshold where the axion mass surpasses the Hubble expansion, marked as $m_{\rm a} = 3H$. This threshold stands as a vital indicator of the significance of the axion mass concerning the Universe's expansion rate. Beyond this threshold, the axion mass takes precedence over the Hubble expansion, significantly influencing their dynamics. In addition, this panel incorporates $\omega_{\rm a}(k,T)/3H(T)$ for the two modes: $k_{\rm c} = R(T_{\rm c}) \, H(T_{\rm c})$ and $k_{\rm osc} = R(T_{\rm osc}) \, H(T_{\rm osc})$. These modes serve as reference points when assessing the significance of the axion mass concerning the wavenumber.

Our analysis focuses on modes that enter the horizon after $k_{\rm osc}$ due to their crucial role in computing the power spectrum for axion field fluctuations. The choice of this criterion is rooted in the fundamental characteristics of axion dark matter and the specific dynamics associated with the misalignment mechanism. Scales that enter the horizon well before $k_{\rm osc}$ correspond to very small astrophysical scales. In that case $k/R \gg m_{\rm a}$  at horizon crossing. The effect of axion mass will only kick in much later, when $k/R$ drops below $m_{\rm a}$, and from then on all modes will oscillate with the same frequency defined by the mass. The scale $k_{\rm osc}$ is intimately connected to the oscillation scale of axions, representing a critical length scale associated with the characteristic behavior of the axion field. By concentrating on modes with wavenumbers $k \lesssim k_{\rm osc}$, we direct our attention to spatial scales larger than the typical axion oscillation length. The significance of these modes arises from their proximity to the critical threshold where the axion mass becomes dominant over the Hubble expansion. This enables these modes to exert a substantial influence on large-scale structures, increasing the potential for leaving distinct imprints for observations. This deliberate focus is particularly crucial for understanding the misalignment mechanism, where axions are produced through the oscillations of a field initially displaced from its minimum. These chosen modes play a pivotal role in shaping the evolution of axion density perturbations. We can also ignore modes that enter the horizon at $T \ll T_{\rm c}$, as they are frozen superhorizon modes during the onset of the axion 
mass and they do not take notice of the effect of the onset of the axion mass. For their evolution we can just assume  $m_{\rm a} = m_{\rm 0}$ at all times. Thus we can restrict our analysis to $k_{\rm c} < k < k_{\rm osc}$.

While the step-function model in Eq.~\eqref{eq:2.7} is a simplification, it serves as a reliable approximation for capturing the essential dynamics of axion mass acquisition. As illustrated in the left panel of Figure~\ref{fig:mass}, the actual temperature dependence of $m_{\rm a}(T)$, derived from lattice QCD data, exhibits a rapid rise near the QCD crossover temperature $T_{\rm c}$, which motivates modeling the mass as turning on instantaneously around a characteristic temperature. Importantly, the sharpness of this transition becomes more pronounced at higher values of $f_a$, where axion oscillations commence later and the relevant mass scale is lower, further compressing the temperature interval over which the mass turns on. Therefore, for $f_a \gtrsim 10^{12}\, \text{GeV}$, which corresponds to the regime of primary interest in this work, the step-function approximation captures the relevant dynamics with sufficient accuracy. In the following analysis, we focus on the case $f_a = 10^{16}\, \text{GeV}$, where the approximation remains particularly well justified and offers a practical simplification for studying the generation of axion field fluctuations. We have also explicitly compared this approximation to the full temperature-dependent mass and found good agreement in the regimes relevant for the evolution of field perturbations (see Figure~\ref{fig:mass}). For the purposes of computing the power spectrum of axion fluctuations in Section~\ref{sec:4}, this approach allows for analytical tractability without compromising the physical accuracy of our conclusions.

\section{Equation of motion for the axion field} \label{sec:3}

In this section, we analyze the dynamics of the axion field $a \equiv a(\vb{x},t)$ within an expanding Universe. It is often convenient to express this field in terms of the dimensionless misalignment field $\theta(\vb{x},t)=a(\vb{x},t)/f_{\rm a}$. The action for the misalignment field is given by
\begin{equation} \label{eq:3.1}
S_{\rm \theta} = \int d^4 x \sqrt{-g} f_{\rm a}^2 \left( - \frac{1}{2} g^{\mu\nu} \partial_{\rm \mu} \theta \partial_{\rm \nu} \theta  - V(\theta) \right) \,,
\end{equation}
where $g$ is the determinant of the spacetime metric $g_{\mu \nu}$, and $V(\theta)$ represents the potential energy density associated with the misalignment field. As detailed in~\cite{Gross:1980br}, this potential can be expressed as
\begin{equation} \label{eq:3.2}
V(\theta) = m_{\rm a}^2(t) \left( 1- \cos \theta \right) \simeq \frac{1}{2} m_{\rm a}^2(t) \theta^2 \,.
\end{equation}
For small values of $\theta$, the harmonic approximation of the axion potential is employed. 
The equation of motion for the misalignment field can be derived from the action in Eq.~\eqref{eq:3.1} using the principle of least action, $\delta S_{\rm \theta}/\delta \theta=0$. This leads to the Klein-Gordon equation
\begin{equation} \label{eq:3.3}
- \frac{\partial_\nu \left( \sqrt{-g} g^{\mu \nu} \partial_\mu \theta \right)}{\sqrt{-g}} + \frac{\partial V}{\partial \theta} = 0\, .
\end{equation}
The equation of motion is simplified considerably, when we look at its homogeneous and isotropic solutions. Therefore, we use the concept of cosmological perturbations to expand the axion field around its spatial mean, expand it in Fourier modes and study their evolution in the following. The corresponding wavelength of each comoving mode $k$ scales as $\lambda = 2\pi R/k$.  

\subsection{Perturbations in the axion field} \label{sec:3.1}

In order to describe these cosmological perturbations, we consider small deviations from the Friedmann-Lemaître-Robertson-Walker (FLRW) metric. In longitudinal gauge, and in the absence of anisotropic stress, the perturbed metric in a flat FLRW spacetime can be expressed as~\cite{Bardeen:1980kt, Mukhanov:1990me}
\begin{equation} \label{eq:3.5}
{\rm d}s^2 = -(1 + 2\Phi) {\rm d}t^2 + R^2(t) (1 - 2\Phi) {\rm d}\vb{x}^2 \,,
\end{equation}
where $\Phi$ represents the gravitational metric potential. Further, we also introduce an arbitrary fluctuation in the axion field around its background value
\begin{equation} \label{eq:3.7}
\theta (\vb{x},t)= \bar{\theta}(t) + \delta \theta (\vb{x},t) \,.
\end{equation}
In the first step, this allows us to obtain an equation of motion for the 
mean misalignment field
\begin{equation} \label{eq:3.3}
\dfrac{{\rm d}^2 \bar\theta}{{\rm d}t^2} + 3 H \dfrac{{\rm d}\bar\theta}{{\rm d}t} + \dfrac{\partial V }{\partial \theta}(\bar \theta) = 0 \,.
\end{equation}
Since the misalignment potential $V$ depends on the temperature of the QCD matter, we also consider adiabatic temperature fluctuations
\begin{equation} \label{eq:3.8}
T(\vb{x},t) = \bar{T}(t) + \delta T(\vb{x},t) \,, 
\end{equation}
where $\bar{T}$ represents the background temperature and $\delta T$ represents the temperature fluctuation around $\bar{T}$. Considering the harmonic approximation of the potential Eq.~\eqref{eq:3.2}, and taking its derivative with respect to $\theta$, we obtain at linear order
\begin{equation} \label{eq:3.9}
\frac{\partial V}{\partial \theta} \simeq m_{\rm a}^2(\bar{T}) \bar{\theta} + m_{\rm a}^2(\bar{T}) \delta \theta + \frac{{\rm d} m_{\rm a}^2}{{\rm d} T}(\bar{T}) \bar{\theta} \delta T \,.
\end{equation}
The homogeneous equation of motion is thus given by
\begin{equation} \label{eq:3.10}
\dfrac{{\rm d}^2 \bar{\theta}}{{\rm d}t^2} + 3H(t) \dfrac{{\rm d} \bar{\theta}}{{\rm d}t} + m_{\rm a}^2(\bar{T}) \bar{\theta} = 0 \,.
\end{equation}
The equation of motion for the inhomogeneous perturbations of the misalignment field, restricted to first-order perturbations, is given by
\begin{equation} \label{eq:3.6}
\dfrac{\partial^2 \delta\theta}{\partial t^2} + 
3H(t) \dfrac{\partial \delta \theta}{\partial t} - 
\frac{1}{R^2(t)} \nabla^2 \delta \theta + 
m_{\rm a}^2(\bar T) \delta \theta 
= 
- \frac{{\rm d} m_{\rm a}^2}{{\rm d} T}(\bar{T}) \bar{\theta} \delta T
+ 2 m_{\rm a}^2(\bar{T}) \bar{\theta} \Phi + 
4  \dfrac{{\rm d}\bar \theta}{{\rm d} t} \dfrac{\partial \Phi}{\partial t} \,. 
\end{equation}



To simplify the analysis, we introduce a Fourier transformation with respect to the spatial coordinates and express the misalignment field as
\begin{equation} \label{eq:3.12}
\vartheta(\vb{k},t) = \int d^3 x \, e^{-i\vb{k} \vb{x}}  \delta \theta(\vb{x},t) \,.
\end{equation}
The corresponding Fourier transforms of $\delta T$ and $\Phi$ are denoted by $\mathcal{T}$ and $\varphi$, respectively.
Applying this Fourier transformation, the perturbed equation of motion~\eqref{eq:3.14} for single $k$ modes can be written as
\begin{equation}
\dfrac{{\rm d}^2 \vartheta}{{\rm d}t^2} + 
3H(t) \dfrac{{\rm d}\vartheta}{{\rm d}t} + \frac{k^2}{R^2(t)} \vartheta + m_{\rm a}^2(\bar{T}) \vartheta = - \frac{{\rm d} m_{\rm a}^2}{{\rm d} T}(\bar{T}) \bar{\theta} \mathcal{T}
+ 2  m_{\rm a}^2(\bar{T}) \bar{\theta} \varphi + 
4  \dfrac{{\rm d}\bar \theta}{{\rm d} t} \dfrac{{\rm d} \varphi}{{\rm d} t} \,. 
\label{eq:modes}
\end{equation}
In this equation, several terms can be identified that couple the axion field to the dominant plasma. The terms containing $m_{\rm a}$ represent the coupling due to the temperature-dependent axion mass. The terms involving $\varphi$ describe gravitational couplings between the axion field perturbations and the surrounding gravitational potential. During the radiation-dominated epoch, the gravitational potential $\varphi$ is primarily determined by the plasma and its acoustic oscillations. For superhorizon modes, the term involving ${\rm d} \varphi/{\rm d} t$ can be neglected, as $\varphi$ is approximately constant on those scales. This is valid for the modes of interest, specifically those where $k_{\rm osc} < k < k_{\rm c}$, prior to the critical temperature $\bar{T}_{\rm c}$. The term containing $m_{\rm a}^2 \bar{\theta}$ introduces small corrections, which are typically negligible compared to $m_{\rm a}^2$, as $\bar{\theta} \ll 1$ in the pre-inflationary scenario while $\vartheta$ and $\varphi$ are expected to be of the same order of magnitude. 
Therefore, we can safely drop the last two terms in (\ref{eq:modes}). 
The term involving temperature perturbations, $({\rm d} m_{\rm a}^2/{\rm d} T) \bar{\theta} \mathcal{T}$, can potentially drive oscillations in the axion field, especially for a sudden onset of the axion mass. The significance of this term depends on the specific behavior of $m_{\rm a}(\bar{T})$ and the magnitude of the temperature fluctuations $\delta T$. 
\begin{equation} \label{eq:3.14}
\dfrac{{\rm d}^2 \vartheta}{{\rm d}t^2} +
3H(t) \dfrac{{\rm d} \vartheta}{{\rm d}t} +
\left(\frac{k^2}{R^2(t)} + m_{\rm a}^2(\bar{T}) \right) \vartheta = 
-  \frac{{\rm d} m_{\rm a}^2}{{\rm d} T}(\bar{T}) \bar{\theta} \mathcal{T} \,.
\end{equation}
This equation enables us to explore the complex interplay between axion field dynamics and the surrounding environment. The left-hand side corresponds to the first-order perturbations in the axion field, capturing deviations from its homogeneous state and their effects on the field's evolution. Conversely, the right-hand side represents the perturbation contribution induced by adiabatic temperature fluctuations, resulting from the coupling with the surrounding plasma. By studying this interplay, we gain insights into the formation and evolution of structures within the axion field, such as density fluctuations and the generation of power spectra. This equation facilitates the analysis of how background dynamics interact with perturbations, providing a valuable understanding of the axion field's behavior.


\subsection{Evolution of the background axion field} \label{sec:3.2}

For numerical studies of the axion evolution in the radiation dominated epoch of the Universe, it is convenient to express the equation of motion as a function of temperature instead of time, see also Eq.~\eqref{eq:2.3}. This transformation yields the following form for the unperturbed equation of motion~\eqref{eq:3.10}
\begin{equation} \label{eq:3.4}
\dfrac{{\rm d}^2 \bar{\theta}}{{\rm d}T^2} + \left[ 3H(\bar{T}) \dfrac{{\rm d}t}{{\rm d}T} - \dfrac{{\rm d}^2 t}{{\rm d}T^2} / \dfrac{{\rm d}t}{{\rm d}T}  \right] \dfrac{{\rm d}\bar{\theta}}{{\rm d}T}+ m_{\rm a}^2(\bar{T}) \bar{\theta} \left( \dfrac{{\rm d}t}{{\rm d}T} \right)^2= 0 \,.
\end{equation}
For a radiation dominated Universe with $c_s^2 = 1/3$ and $H \propto \bar{T}^2$, the term in the square bracket vanishes and it becomes nontrivial in the vicinity of the QCD transition at $\bar{T}_{\rm c}$, when $c_{\rm s}^2 < 1/3$ and $H \propto \sqrt{g(\bar{T})}\bar{T}^2$. The evolution of the background equation~\eqref{eq:3.4} can be understood as follows. Initially, at high temperatures $\bar{T} \gg \bar{T}_{\rm c}$ the axion is effectively massless and only the constant term $\bar{\theta}(\bar{T}) = \theta_{\rm i}$ is relevant.
In this 
regime, the field is frozen at its initial value $\theta_{\rm i}$, fixed during cosmological inflation due to Hubble friction, as can be seen from the structure of Eq.~\refeq{eq:3.3}. 

As the temperature decreases to around the QCD phase transition temperature $\bar{T} \sim \bar{T}_{\rm c}$, the axion mass becomes relevant, and the axion field starts to oscillate at $\bar{T}_{\rm osc}$. 
After the axion mass is switched on, the last term in Eq.~\eqref{eq:3.4} becomes effectively a temperature dependent mass term $\propto \bar{T}^{-6}$. To ensure a smooth analytic description from the frozen axion field an  oscillating one, we can make the  ansatz
\begin{equation} \label{eq:3.15}
\bar{\theta}(\bar{T}) = 
\theta_{\rm i}\left( 1 - A \left[ \frac{\bar{T}_{\rm osc}}{\bar{T}} \right]^\alpha \right)\,, \quad {\rm for} \quad \bar{T} \gtrsim \bar{T}_{\rm osc} \,,
\end{equation}
where $A = 9/20$ and $\alpha = 4$ are obtained solving for the leading terms in \eqref{eq:3.4}.

However, such a polynomial ansatz cannot describe the subsequent oscillations. To simplify the analysis and obtain an approximate solution for the oscillatory phase, the  Wentzel-Kramers-Brillouin (WKB) approximation is commonly used. Within this approximation, we apply the ansatz $\theta(\bar{T}) = A(\bar{T}) \exp({i \phi(\bar{T})})$, where $A(\bar{T})$ and $\phi(\bar{T})$ are slowly varying functions of time. By substituting this ansatz into the equation of motion, we can derive an approximate solution for the axion field at $\bar{T} \ll \bar{T}_{\rm c}$. Therefore, the axion field within the WKB approximation within a perfectly radiation dominated phase with $c_{\rm s}^2 = 1/3$ and $H \propto \bar{T}^2$ reads 
\begin{equation} \label{eq:3.16}
\bar{\theta}(\bar{T}) = B \left[ \frac{\bar{T}}{\bar{T}_{\rm osc}}\right]^{3/2} \cos\left(\frac{3}{2} \frac{\bar{T}_{\rm osc}^2}{\bar{T}^2} + \beta \right) \,, \quad {\rm for} \quad \bar{T} \lesssim \bar{T}_{\rm osc} \,.
\end{equation}
Here, $B$ and $\beta$, the amplitude and phase can be obtained from the numerical solution of Eq.~\eqref{eq:3.10}. 

\begin{figure}[t!]
\centering
\includegraphics[width=0.456596\textwidth]{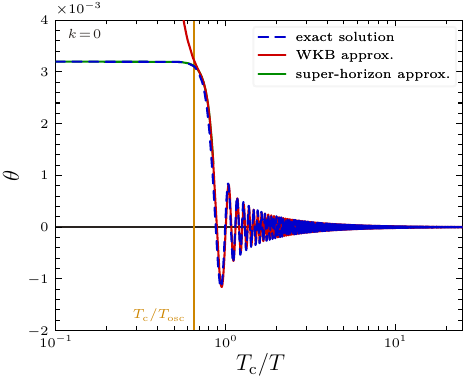}
\includegraphics[width=0.536004\textwidth]{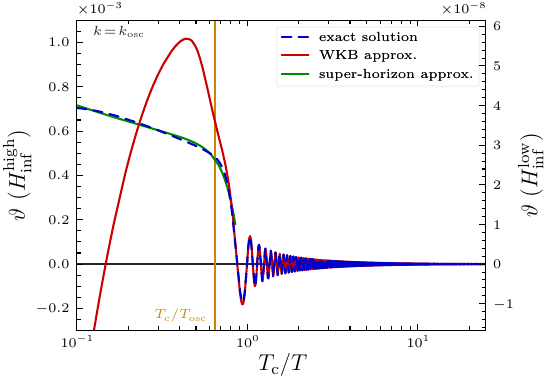}
\caption{The left panel shows the evolution of the background axion field $\bar{\theta}$ with respect to temperature $T$. The numerical solution in blue obtained by solving Eq.~\eqref{eq:3.10} is shown along with the approximate solution in red obtained using Eqs.~\eqref{eq:3.15} and~\eqref{eq:3.16}. The right panel shows the evolution of the axion field fluctuations $\vartheta$ with respect to temperature $T$ for the mode with $k = k_{\rm osc}$, within the $H_{\rm inf}^{\rm high} = 4.44 \times 10^{13} \, {\rm GeV}$ and $H_{\rm inf}^{\rm low} = 2.48 \times 10^9 \, {\rm GeV}$ regimes, as indecated by the left and right vertical axes labels, respectively. The numerical solution in blue obtained by solving Eq.~\eqref{eq:3.17} is shown along with the approximate solution in red obtained using Eq.~\eqref{eq:3.18}. The PQ scale is set to $f_{\rm a}= 10^{16} \, {\rm GeV}$.}
\label{fig:f0}
\end{figure}

The left panel of Figure~\ref{fig:f0} illustrates the evolution of the background axion field $\bar{\theta}$ obtained numerically by solving Eq.~\eqref{eq:3.10}, along with the approximate solution given by Eqs.~\eqref{eq:3.15} and~\eqref{eq:3.16}. The initial value of the amplitude for the field is set to $\theta_{\rm i} \approx 3.2 \times 10^{-3}$~\cite{Sakharov:2021dim}. This initial value is chosen deliberately, as it aligns with the potential to elucidate the overall dark matter abundance in the Universe, as mentioned previously. By comparing the numerical and approximate solutions, we can observe the oscillatory behavior of the field as the temperature evolves. The amplitude of the oscillations decreases with time due to the expansion of the Universe, while the phase of the oscillations advances. This evolution of the background field sets the foundation for the study of perturbations and their impact on the formation of structures in the axion field in the following sections.

\subsection{Evolution of the axion field perturbations} \label{sec:3.3}

Similarly, the temperature-dependent form of the perturbed equation of motion~\eqref{eq:3.14} can be written as
\begin{equation} \label{eq:3.142}
\begin{aligned}
\dfrac{{\rm d}^2 \vartheta}{{\rm d}T^2} &+ \left(3H(\bar{T}) \dfrac{{\rm d}t}{{\rm d}T} - \dfrac{{\rm d}^2 t}{{\rm d}T^2} / \dfrac{{\rm d}t}{{\rm d}T}  \right) \dfrac{{\rm d} \vartheta}{{\rm d}T} + \left(\frac{k^2}{R^2(\bar{T})} + m_{\rm a}^2(\bar{T}) \right) \left( \dfrac{{\rm d}t}{{\rm d}T} \right)^2 \vartheta \\&= -  \frac{\partial m_{\rm a}^2(\bar{T})}{\partial T} \left( \dfrac{{\rm d}t}{{\rm d}T} \right)^2 \bar{\theta} \mathcal{T} \,.
\end{aligned}
\end{equation}
The evolution of perturbations within the axion field can be thoroughly examined by considering this equation. The LHS of this equation accounts for the field fluctuations, while the RHS accounts for the adiabatic temperature fluctuations caused by primordial plasma temperature fluctuations. For the moment, let us neglect the adiabatic temperature fluctuations and therefore we arrive at the simplified perturbed equation of motion for the field fluctuations seeded by quantum fluctuations during inflation
\begin{equation} \label{eq:3.17}
\dfrac{{\rm d}^2 \vartheta}{{\rm d}T^2} + \left(3H(\bar{T}) \dfrac{{\rm d}t}{{\rm d}T} - \dfrac{{\rm d}^2 t}{{\rm d}T^2} / \dfrac{{\rm d}t}{{\rm d}T}  \right) \dfrac{{\rm d} \vartheta}{{\rm d}T} + \left(\frac{k^2}{R^2(\bar{T})} + m_{\rm a}^2(\bar{T}) \right) \left( \dfrac{{\rm d}t}{{\rm d}T} \right)^2 \vartheta = 0 \,.
\end{equation}
This equation can be solved numerically for individual $k$ modes, revealing that the behavior of these solutions closely mirrors that of the background solutions. Notably, the amplitude of the fluctuations diminishes as $k$ increases. This observation allows us to propose the following approximate solution for the axion field fluctuations
\begin{equation} \label{eq:3.18}
\vartheta(\bar{T}) = 
\begin{dcases}
\vartheta_{\rm i}\left(1 - C \left[\frac{\bar{T}}{\bar{T}_{\rm osc}} \right]^\gamma \right) \,, 
\quad & {\rm for} \quad \bar{T} \gtrsim \bar{T}_{\rm osc} \,, \\
D 
\sqrt{\frac{\omega_{\rm a}(\bar{T}_{\rm osc})}{\omega_{\rm a}(\bar{T})}} \, \left[ \frac{R(\bar{T}_{\rm osc})}{R(\bar{T})} \right]^{3/2} \, \cos\left( \int \frac{\omega_{\rm a}(\bar{T})}{H(\bar{T}) \bar{T}} \, {\rm d}T + \zeta \right) \,, \quad & {\rm for} \quad \bar{T} \lesssim \bar{T}_{\rm osc} \,,
\end{dcases}
\end{equation}
where $\vartheta_{\rm i}$ denotes the initial value of the axion field fluctuations, and $C$ and $\gamma$ can be obtained from the numerical solution to ensure the approximate ansatz fits the numerical solution accurately. Here, $\omega_{\rm a}(\bar{T}) = \sqrt{k^2/R^2(\bar{T}) + m_{\rm a}^2(\bar{T})}$, and $D$ and $\zeta$ are the amplitude and phase, which can be obtained from the numerical solution of Eq.~\eqref{eq:3.17}. 

The typical amplitude of initial quantum fluctuations is of the order of $\vartheta_{\rm i} \sim H_{\rm inf}/(2\pi f_{\rm a})$, where $H_{\rm inf}$ is the Hubble parameter during inflation~\cite{Linde:1991km, Lyth:1992tx, Beltran:2006sq, Lyth:1992tw, Sakharov:2021dim}. To estimate the value of $\vartheta_{\rm i}$ at the end of inflation, we need to determine the inflationary energy scale. In single-field slow-roll inflation, the expansion rate during inflation can be expressed as~\cite{Liddle:2000cg}
\begin{equation} \label{eq:3.19}
H_{\rm inf} \approx \pi M_{\rm Pl} \sqrt{\frac{r A_{\rm s}}{2}} \,,
\end{equation}
where $r$ is the tensor-to-scalar ratio, $A_{\rm s} \approx 2.105(30) \times 10^{-9}$ is the amplitude of scalar perturbations~\cite{Planck:2018vyg}, and $M_{\rm Pl} \simeq 2.435 \times 10^{18} \, {\rm GeV}$ is the reduced Planck mass.

The upper limit $r < 0.056$ set by Planck and BICEP2 observations~\cite{Planck:2018jri, BICEP2:2018kqh} was recently improved to $r < 0.032$, as reported in~\cite{Tristram:2021tvh}. This constraint translates into an upper bound on the Hubble parameter during high-energy scale inflation as $H_{\rm inf}^{\rm high} < 4.44 \times 10^{13} \, {\rm GeV}$. Assuming $f_{\rm a}= 10^{16} \, {\rm GeV}$, the corresponding amplitude of initial quantum fluctuations is of the order $\vartheta_{\rm i} \sim 7 \times 10^{-4}$. 
When considering low-energy scale inflation, motivated by models such as K\"ahler-moduli inflation~\cite{Conlon:2005jm}, it is characterized by a significantly reduced tensor-to-scalar ratio, $r \sim 10^{-10}$. This corresponds to a Hubble parameter of $H_{\rm inf}^{\rm low}  < 2.48 \times 10^9 \, {\rm GeV}$. Assuming $f_{\rm a}= 10^{16} \, {\rm GeV}$, this results in an amplitude of quantum fluctuations of the order $\vartheta_{\rm i} \sim 4 \times 10^{-8}$. 


In the pre-inflationary scenario, the axion field remains energetically subdominant to the cosmic expansion, with its energy density perturbations primarily arising from quantum fluctuations generated during inflation. In the high-energy inflationary regime, where $H_{\rm inf}$ is close to the PQ scale $f_{\rm a}$, the amplitude of these fluctuations are suppressed but within the observational bounds, consequently preventing the formation of significant primordial CDM isocurvature perturbations~\cite{Tenkanen:2019xzn, BICEP:2021xfz, Planck:2018jri, Planck:2018vyg}. Conversely, in the low-energy inflationary regime, where $H_{\rm inf}$ is relatively small compared to $f_{\rm a}$, the amplitude of quantum fluctuations are further suppressed, significantly reducing isocurvature perturbations, in agreement with the current Planck observations~\cite{Planck:2018jri, Planck:2018vyg}. For our purposes, the low-energy inflationary regime is of particular interest, as adiabatic temperature fluctuations dominate energy density perturbations compared to quantum fluctuations, as we will discuss in the next sections.

The right panel of Figure~\ref{fig:f0} depicts the evolution of axion field fluctuations $\vartheta$ as a function of temperature $T$ for the mode with $k = k_{\rm osc}$. The panel presents numerical results for the two different inflationary Hubble parameter regimes: $H_{\rm inf}^{\rm high} = 4.44 \times 10^{13} \, \text{GeV}$ and $H_{\rm inf}^{\rm low} = 2.48 \times 10^9 \, \text{GeV}$, which are represented by distinct scales on the $y$-axis. The numerical solution in blue, obtained by solving Eq.~\eqref{eq:3.17}, is shown along with the approximate solution in red, obtained using Eq.~\eqref{eq:3.18}. By comparing the two solutions, we observe the oscillatory behavior of the fluctuations as the temperature changes. The amplitude of the fluctuations decreases over time due to the expansion of the Universe, similar to the behavior of the background field. Additionally, the phase of the fluctuations advances with time. 

For simplicity of notation, we drop the bars from $\theta$ and $T$ in the following, assuming they represent the background values unless stated otherwise.

\section{Revising the perturbed equation of motion} \label{sec:4}

To proceed with our analysis, we return back to the simplified axion mass model introduced in Section~\ref{Sec:2.3}, which is governed by Eq.~\eqref{eq:2.7}. This model captures the rapid transition in axion mass at the temperature $T_{\rm osc}$, at which axion field oscillations commence. Within this framework, the temperature dependence of the axion mass is described by
\begin{equation} \label{eq:4.1}
\frac{\partial m_{\rm a}^2(T)}{\partial T} = - m_{\rm a}^2 \, \delta_{\rm D} (T_{\rm osc} - T) \,,
\end{equation}
where $\delta_{\rm D}(x)$ is the Dirac delta function. This greatly simplifies the treatment of the temperature-dependent axion mass. Applying this approximation, the perturbed equation of motion, as given in Eq.~\eqref{eq:3.142}, which governs the evolution of axion field perturbations, becomes
\begin{equation} \label{eq:4.2}
\begin{aligned}
\dfrac{{\rm d}^2 \vartheta}{{\rm d}T^2} &+ \left(3H(T) \dfrac{{\rm d}t}{{\rm d}T} - \dfrac{{\rm d}^2 t}{{\rm d}T^2} / \dfrac{{\rm d}t}{{\rm d}T}  \right) \dfrac{{\rm d} \vartheta}{{\rm d}T} + \left(\frac{k^2}{R^2(T)} + m_{\rm a}^2(T) \right) \left( \dfrac{{\rm d}t}{{\rm d}T} \right)^2 \vartheta \\&= m_{\rm a}^2 \, \delta_{\rm D} (T_{\rm osc} - T) \left( \dfrac{{\rm d}t}{{\rm d}T} \right)^2 \theta(T_{\rm osc}) \mathcal{T} \,.
\end{aligned}
\end{equation}
To analyze the behavior of the axion field in the presence of adiabatic temperature fluctuations, we consider two regimes: (i) $T \gtrsim T_{\rm osc}$, before the mass transition, and (ii) $T \lesssim T_{\rm osc}$, after the mass transition.

In the first case, when $T \gtrsim T_{\rm osc}$, the axions are massless. Assuming that the previously discussed quantum fluctuations are small, the axion field remains homogeneous. This leads to the trivial solution
\begin{equation} \label{eq:4.3}
\vartheta (\vb{k}, T)=0 \,, \quad {\rm for} \quad T>T_{\rm osc} \,.
\end{equation}

In the second case, when $T \lesssim T_{\rm osc}$, the axions acquire a nonzero mass $m_{\rm a}$, and perturbations are expected to evolve. 
In our approach, we describe this evolution using the same equation as Eq.~\eqref{eq:3.17}, which also governs quantum-induced fluctuations but with a different mechanism for the initial conditions.
The key distinction lies in the fact that the presence of adiabatic temperature fluctuations does not explicitly appear in the equation itself. Instead, their effect is encoded through the initial conditions. By imposing continuity conditions on the axion field and its derivative across $T_{\rm osc}$, the information from the adiabatic fluctuations is effectively captured in the evolution of the perturbations. This approach allows us to incorporate the influence of adiabatic temperature fluctuationsafter  without altering the form of the equation.

At $T=T_{\rm osc}$, the axion mass undergoes an abrupt change. To comprehensively analyze this transition, it is crucial to account for both the field and its derivative. Despite the sudden shift in mass, continuity in the field is essential for ensuring a seamless evolution. This demands that the field maintains the same value just before and after the transition. 
Such continuity around the transition moment is crucial for maintaining the coherence of the system and can be expressed as
\begin{equation} \label{eq:4.5}
[\vartheta] = \lim_{\rm \epsilon \to 0} \vartheta \big\vert_{T_{\rm osc} + \epsilon}^{T_{\rm osc} - \epsilon} =  \lim_{\rm \epsilon \to 0} \int_{T_{\rm osc} + \epsilon}^{T_{\rm osc} - \epsilon} {\rm d}T \, \vartheta' \,,
\end{equation}
where primes denote derivatives with respect to $T$, $\vartheta(T_{\rm osc} + \epsilon)$ represents the value of the perturbed field just above the transition at $T_{\rm osc}$, and $\vartheta(T_{\rm osc} - \epsilon)$ represents the value of the perturbed field just below the transition. Similarly, we can define the continuity condition for the derivative of the field
\begin{equation} \label{eq:4.6}
[\vartheta'] = \lim_{\rm \epsilon \to 0} \vartheta' \big\vert_{T_{\rm osc} + \epsilon}^{T_{\rm osc} - \epsilon} =  \lim_{\rm \epsilon \to 0} \int_{T_{\rm osc} + \epsilon}^{T_{\rm osc} - \epsilon} {\rm d}T \, \vartheta'' \,.
\end{equation}
Applying Eq.~\eqref{eq:4.5} to the perturbed equation of motion~\eqref{eq:4.2} and then substituting into Eq.~\eqref{eq:4.6}, we obtain
\begin{equation} \label{eq:4.7}
\begin{aligned}
{[}\vartheta'] &= \lim_{\rm \epsilon \to 0} \int_{T_{\rm osc} + \epsilon}^{T_{\rm osc} - \epsilon} {\rm d}T \, m_{\rm a}^2 \delta_{\rm D} (T_{\rm osc}-T) \theta(T_{\rm osc}) \mathcal{T} \\
&= - m_{\rm a}^2 \left( \left. \dfrac{{\rm d}t}{{\rm d}T} \right|_{T_{\rm osc}} \right)^2 \theta(T_{\rm osc}) \mathcal{T} \\
&= - m_{\rm a}^2 \left( \left. \dfrac{{\rm d}t}{{\rm d}T} \right|_{T_{\rm osc}} \right)^2 \theta(T_{\rm osc}) T_{\rm osc}  A_{\rm k} \cos \left[ c_{\rm s}(T_{\rm osc}) k \tau(T_{\rm osc}) \right] \,.
\end{aligned}
\end{equation}
The term $\theta(T_{\rm osc})$ represents the homogeneous background solution at the oscillation temperature $T_{\rm osc}$. To arrive at this step, we use the expression for temperature fluctuations during the radiation-dominated era, which can be written as
\begin{equation} \label{eq:4.8}
\frac{\mathcal{T}}{T} = \frac{\delta \rho(T)}{3\left(\rho(T)+p(T)\right)} = A_{\rm k} \cos \left[ c_{\rm s}(T) k \tau(T) \right] \,,
\end{equation}
where $\rho(T)$ represents the energy density, $p(T)$ indicates the pressure, $A_{\rm k} \sim \sqrt{A_{\rm s}}$ is the amplitude of the temperature fluctuations at wavenumber $k$, 
and $\tau(T) \equiv \int {\rm d}T ({\rm d}t/{\rm d}T)/R(T)$ denotes the conformal time.

Thus, to solve the perturbed equation of motion~\eqref{eq:4.2} for $T \leq T_{\rm osc}$, we only need to consider the simplified perturbed equation of motion~\eqref{eq:3.17}
with the initial conditions
\begin{equation}  \label{eq:4.10}
\begin{aligned}
\vartheta (\vb{k}, T_{\rm osc}) &= 0 \,, \\
\vartheta' (\vb{k}, T_{\rm osc}) &= - m_{\rm a}^2 \left( \left. \dfrac{{\rm d}t}{{\rm d}T} \right|_{T_{\rm osc}} \right)^2 \theta(T_{\rm osc}) T_{\rm osc} A_{\rm k} \cos \left[ c_{\rm s}(T_{\rm osc}) k \tau(T_{\rm osc}) \right] \,.
\end{aligned}
\end{equation}
Using these initial conditions, we can solve the simplified perturbed equation of motion~\eqref{eq:3.17} numerically to investigate the evolution of axion field perturbations in the presence of adiabatic temperature fluctuations. This comprehensive approach allows us to understand the combined effects of inflationary quantum field fluctuations and adiabatic temperature fluctuations on the behavior of the axion field in the early Universe. From the last equation, and using the approximation $c_{\rm s}(T) = c/\sqrt{3}$ for the sound speed in the radiation-dominated era (where $c=1$ denotes the speed of light), we observe that the solution depends on the term $m_{\rm a}^2 \theta(T_{\rm osc})/H(T_{\rm osc})$. This dependency indicates that the inhomogeneity decays with time. 

\begin{figure}[t!]
\centering
\includegraphics[width=0.536004\textwidth]{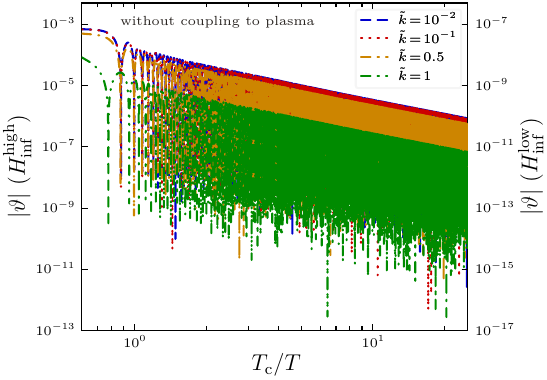}
\includegraphics[width=0.456596\textwidth]{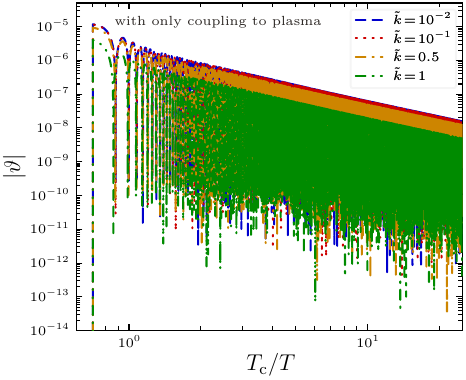}
\caption{The left panel shows the evolution of the axion field for different modes with $\tilde{k}=k/k_{\rm osc}$, seeded by inflationary quantum field fluctuations with $H_{\rm inf}^{\rm high} = 4.44 \times 10^{13} \, {\rm GeV}$ and $H_{\rm inf}^{\rm low} = 2.48 \times 10^9 \, {\rm GeV}$, respectively (left and right vertical axes labels). The right panel shows the evolution of the perturbation in the axion misalignment field for different modes $\tilde{k}$ induced by adiabatic temperature fluctuations during the onset of the axion mass. The PQ scale is set to $f_{\rm a}= 10^{16} \, {\rm GeV}$.}
\label{fig:fd0}
\end{figure} 

The left panel of Figure~\ref{fig:fd0} illustrates the evolution of the axion field for different $k$ modes, considering only inflationary quantum field fluctuations within the high-energy regime ($H_{\rm inf}^{\rm{high}} = 4.44 \times 10^{13}$ GeV) and low-energy regime ($H_{\rm inf}^{\rm{low}} = 2.48 < 10^9$ GeV). The right panel of Figure~\ref{fig:fd0} displays the evolution of the perturbations in the axion field for different $k$ modes, considering the induced adiabatic temperature fluctuations resulting from the coupling with the surrounding plasma. This plot highlights how quantum fluctuations lead to oscillations in the axion field as the Universe expands, with the amplitude of these oscillations decreasing as the Universe cools down. By comparing both panels, we can discern the relative contributions of quantum fluctuations and induced adiabatic perturbations to the overall dynamics of the axion field. These visualizations provide valuable insights into the behavior of the axion field under different initial conditions and perturbative influences, illustrating the distribution of the contribution of different $k$ modes to the field perturbations as discussed in the previous section. Here, $\tilde{k} = k/k_{\rm osc}$, where $\tilde{k}$ takes the values $10^{-2}, 10^{-1}, 0.5, 1$. This detailed analysis sheds light on the evolution of perturbations initiated by different sources and their influence on the evolution of structures within the axion density, as we will discuss in the following section.

\section{Energy density distribution and power spectrum} \label{sec:5}

In this section, we examine the energy density distribution and the power spectrum of axion density fluctuations. The background axion energy density $\rho_{\rm a}$ is determined from the solution of the background field equation Eq.~\eqref{eq:3.10}, and is given by  
\begin{equation} \label{eq:5.1}
\bar{\rho}_{\rm a} = f_{\rm a}^2 \left( \frac{1}{2} \dot{\theta}^2 + \frac{1}{2} m_{\rm a}^2 \theta^2 \right) \,.
\end{equation}  
This expression accounts for both the kinetic and potential energy contributions of the field. To study perturbations in the energy density, we employ the simplified perturbed equation of motion Eq.~\eqref{eq:3.17}, where we neglect the interaction with the metric potential and its derivative. The effect of adiabatic temperature fluctuations is incorporated through the initial conditions, as discussed in the previous section. Focusing on modes entering the horizon before $k_{\rm osc}$, we quantify the contributions of individual wavenumber modes $k$ to the total energy density. Specifically, for small axion inhomogeneities, the leading contribution of each $k$ mode, well after the onset of oscillations, is  
\begin{equation} \label{eq:5.2}
\delta \rho_{\rm a}(k) = f_{\rm a}^2 \left( \dot{\theta} \dot{\vartheta}(k) + m_{\rm a}^2 \theta \vartheta(k) \right) \,.
\end{equation}  
Here, $\delta \rho_{\rm a}(k)$ represents the modifications in energy density due to perturbations in both the field's time derivative, $\dot{\vartheta}(k)$, and its value, $\vartheta(k)$. These modes are particularly significant in determining the power spectrum of axion field fluctuations, as they lie near the threshold where the axion mass begins to dominate over the Hubble expansion.

\begin{figure}[t!]
\centering
\includegraphics[width=0.456596\textwidth]{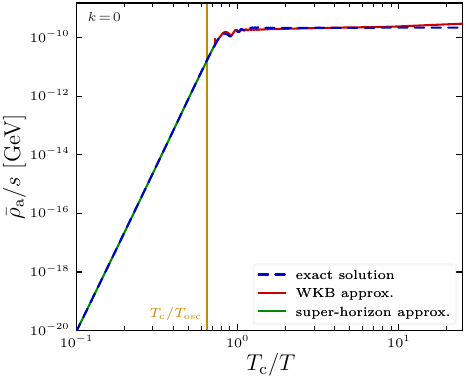}
\includegraphics[width=0.536004\textwidth]{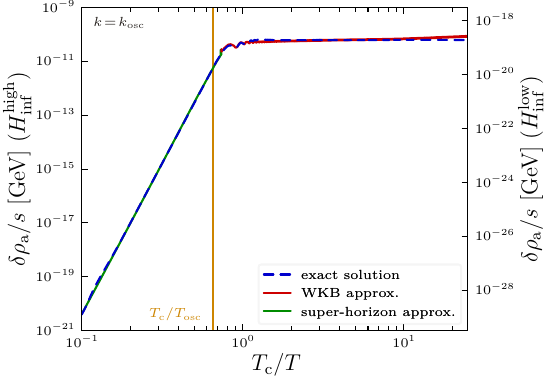}
\caption{Plot showing the evolution of the axion energy density, $\rho_{\rm a}$, and density perturbation, $\delta \rho_{\rm a}$, normalised to the entropy density $s$. The left panel displays the background contribution from the zero-mode with $k=0$. The right panel showcases the contribution from the mode with $k= k_{\rm osc}$, from inflationary quantum field fluctuations for $H_{\rm inf}^{\rm high} = 4.44 \times 10^{13} \, {\rm GeV}$ and $H_{\rm inf}^{\rm low} = 2.48 \times 10^9 \, {\rm GeV}$, respectively (left and right vertical axis labels). The exact numerical result is represented in blue, while the approximate result is depicted in red. The PQ scale is set to $f_{\rm a}= 10^{16} \, {\rm GeV}$. 
}
\label{fig:d0}
\end{figure}

In Figure~\ref{fig:d0}, we present the evolution of the axion energy density, $\rho_{\rm a}$, and the density perturbation, $\delta \rho_{\rm a}$. The left panel corresponds to the background contribution from the zero-mode with $k=0$. The right panel represents the contribution from a specific mode with $k = k_{\rm osc}$, arising from inflationary quantum field fluctuations for both $H_{\rm inf}^{\rm high} = 4.44 \times 10^{13} \, {\rm GeV}$ and $H_{\rm inf}^{\rm low} = 2.48 \times 10^9 \, {\rm GeV}$, as indicated by the left and right vertical axis labels, respectively. We compare the results obtained from the exact numerical solution (depicted in blue) with those from the approximate solution (depicted in red). 

The fluctuation  in the axion energy density can then be defined as
\begin{equation} \label{eq:5.3}
\delta_{\rm a}(\vb{x}) = \frac{\delta \rho_{\rm a}(\vb{x})}{ \bar{\rho}_{\rm a}}  \,.
\end{equation}
This quantity represents the deviation of the energy density from its spatial average $\bar{\rho}_{\rm a}$. Assuming statistical homogeneity and isotropy, we can define the power spectrum $P(\vb{k})$ of the axion density fluctuations in Fourier space using the two-point correlation function
\begin{equation} \label{eq:5.4}
 \expval{\delta_{\rm a}(\vb{k}) \, \delta_{\rm a}(\vb{k'})} = (2 \pi)^3 \delta_{\rm D}(\vb{k}-\vb{k}') P(\vb{k}) \,,
\end{equation}
where the brackets $\expval{\cdot}$ represent an ensemble average and $\delta_{\rm a}(\vb{k})$ is the Fourier transform of the density fluctuation
\begin{equation} \label{eq:5.5}
\delta_{\rm a}(\vb{k}) = \int d^3 x \, e^{-i \vb{k} \cdot \vb{x} } \delta_{\rm a}(\vb{x}) \,.
\end{equation}
The power spectrum $P(\vb{k})$ 
has the dimensions of a volume, providing insight into the contributions of various modes to the overall fluctuation power.

In our analysis, we obtain the power spectrum numerically by employing Eq.~\eqref{eq:5.4}. This approach allows us to gather precise and intricate details regarding the power distribution among various modes. These numerical results become a reference point for comparison with other analytical or semi-analytical estimates. A straightforward estimation of the power spectrum can be characterized by the expression
\begin{equation} \label{eq:5.6}
P(\vb{k}) \propto \left( \frac{\vert \vartheta \vert}{\vert \theta \vert} \right)^2 \,.
\end{equation}
However, it is crucial to consider that the amplitude of non-zero modes diminishes exponentially with respect to $k$, relative to the initial amplitude of the zero mode. This attenuation effect is characterized by an exponential suppression related to $k$, alongside a cut-off scale, as previously discussed. These insights lead to an alternative estimation of the power spectrum
\begin{equation} \label{eq:5.7}
P(\vb{k}) \simeq \exp\left( - \frac{k}{K} \right) \left( \frac{\vert \vartheta_{\rm i} \vert}{\vert \theta_{\rm i} \vert} \right)^2 \,,
\end{equation}
where $K$ is a cut-off scale, $\bar{\theta}_{\rm i}$ represents the background initial value, and $\vartheta_{\rm i}$ is the perturbation initial value. This supplementary equation more accurately represents the power spectrum for non-zero modes, revealing their suppressed amplitudes in comparison to the zero mode due to their exponential dependence on $k$. Understanding these distinctions offers valuable insights into mode behavior and facilitates effective cross-validation of our computational approach with semi-analytical estimates.

To facilitate meaningful comparisons between different approaches and results, it is common practice to work with the dimensionless power spectrum, denoted as $\Delta^2(\vb{k})$. This transformation is achieved through the relation
\begin{equation} \label{eq:5.8}
\Delta^2(\vb{k}) = \frac{k^3}{2 \pi^2} \, P(\vb{k}) \,.
\end{equation}
The dimensionless power spectrum, $\Delta^2(\vb{k})$, is particularly useful as it quantifies the contribution of perturbations as the variance in the density fluctuations per unit logarithmic interval at wavenumber $k$. 

\section{Results and discussion} \label{sec:6} 

In this section, we present the outcomes of our analysis concerning the impact of adiabatic and quantum fluctuations on the density perturbations of axion dark matter. We employ the exact numerical solution of the background axion field Eq.~\eqref{eq:3.10}, to calculate the background density. For quantum fluctuations in high- and low-energy inflation regimes, we use the simplified perturbed equation of motion Eq.~\eqref{eq:3.17}, in which we ignore the interaction with the metric potential and its derivative. 
We implement the influence of the adiabatic temperature fluctuations via initial conditions, as explained in Section~\ref{sec:4}. Our main focus is on modes entering the horizon before $k_{\rm osc}$. 
These modes play a crucial role in calculating the power spectrum of axion field fluctuations due to their proximity to the critical threshold where the axion mass dominates over the Hubble expansion.

\begin{figure}[t!]
\centering
\includegraphics[width=0.536004\textwidth]{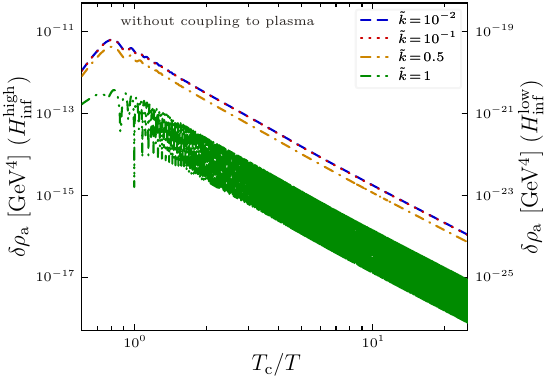}
\includegraphics[width=0.456596\textwidth]{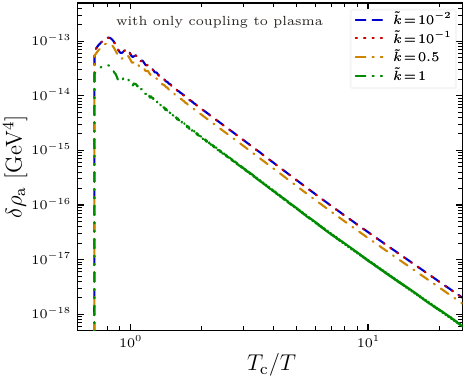}
\caption{The left panel shows the contribution to the axion energy density perturbation, $\delta \rho_{\rm a}$, for different modes with $\tilde{k}=k/k_{\rm osc}$, from inflationary quantum fluctuations for $H_{\rm inf}^{\rm high} = 4.44 \times 10^{13} \, {\rm GeV}$ and $H_{\rm inf}^{\rm low} = 2.48 \times 10^9 \, {\rm GeV}$, as indecated by the left and right vertical axes labels, respectively. The right panel shows the contribution to the axion energy density perturbation, $\delta \rho_{\rm a}$, for different modes with $\tilde{k}=k/k_{\rm osc}$, as induced by adiabatic temperature fluctuations. The PQ scale is set to $f_{\rm a} = 10^{16} \, {\rm GeV}$.}
\label{fig:fd1}
\end{figure}

Initially, we consider the scenario with only inflationary quantum field fluctuations and illustrate them for two different cases of inflationary scales. For a high-energy inflation model with $H_{\rm inf}^{\rm high} = 4.44 \times 10^{13} \, {\rm GeV}$, we apply initial conditions of $\vartheta (\vb{k}, T_{\rm osc}) = 7.07 \times 10^{-4}$ and $\dot{\vartheta}(\vb{k}, T_{\rm osc}) = 0$. This high inflationary scale maximizes the amplitude of the initial quantum fluctuations. For a low-energy inflation model with $H_{\rm inf}^{\rm low} = 2.48 \times 10^9 \, {\rm GeV}$, which accommodate the QCD axion as the total observed CDM, the initial quantum fluctuations are minimized by initial conditions of $\vartheta (\vb{k}, T_{\rm osc}) = 3.95 \times 10^{-8}$ and $\dot{\vartheta}(\vb{k}, T_{\rm osc}) = 0$. These models represent the extreme cases. In both scenarios, we calculate the contribution of each $k$ mode to the energy density. The left panel of Figure~\ref{fig:fd1} illustrates the contribution to the axion energy density perturbation, $\delta \rho_{\rm a}$, for different $k$ modes, assuming only inflationary quantum field fluctuations. 
Here, $\tilde{k} \equiv k/k_{\rm osc}$, where $\tilde{k}$ takes the values $10^{-2}, 10^{-1}, 0.5, 1$.

Subsequently, we incorporate adiabatic temperature fluctuations by considering the initial conditions given by Eq.~\eqref{eq:4.10}. The choice of a low inflationary scale minimizes the influence of quantum fluctuations and highlights the predominant role of adiabatic temperature fluctuations. Similar to the inflationary quantum field fluctuations case, we calculate the contribution of each $k$ mode to the energy density and examine the axion energy density perturbation, $\delta \rho_{\rm a}$. The right panel of Figure~\ref{fig:fd1} illustrates the contribution to the axion energy density perturbation, $\delta \rho_{\rm a}$, for the same 
$\tilde{k}$ modes as above, assuming adiabatic temperature fluctuations. 

\begin{figure}[t!]
\centering
\includegraphics[width=0.4963\textwidth]{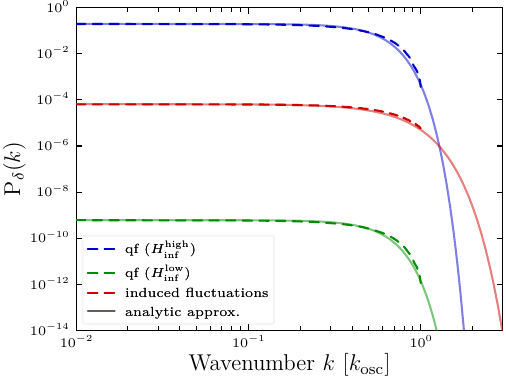}
\includegraphics[width=0.4963\textwidth]{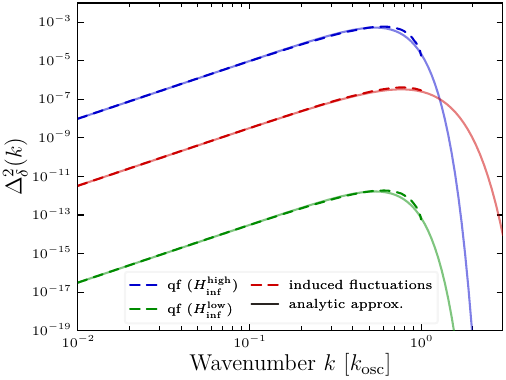}
\caption{The figure shows the power spectrum $P_{\rm \delta}(k)$ and the dimensionless power spectrum $\Delta_{\rm \delta}^2(k)$ of the misalignment axion density fluctuations as a function of the wavenumber $k$. The results obtained assuming only inflationary quantum field fluctuations with $H_{\rm inf}^{\rm high} = 4.44 \times 10^{13} \, {\rm GeV}$ are shown in blue, and with $H_{\rm inf}^{\rm low} = 2.48 \times 10^9 \, {\rm GeV}$ are shown in green, while the results assuming adiabatic temperature fluctuations are represented in red. The dotted lines depict their analytic approximation. The PQ scale is set to $f_{\rm a}= 10^{16} \, {\rm GeV}$.}
\label{fig:dps}
\end{figure}

To further analyze the impact of incorporating adiabatic temperature fluctuations, we investigate the power spectrum $P_{\rm \delta}(k)$ and the dimensionless power spectrum $\Delta_{\rm \delta}^2(k)$ of the misalignment axion density fluctuations as functions of the wavenumber $k$. Figure~\ref{fig:dps} presents the power spectra for several scenarios. The blue curve peaks at $5.95 \times 10^{-4}$ and represents the results obtained assuming only inflationary quantum field fluctuations with high-energy inflation ($H_{\rm inf}^{\rm high} = 4.44 \times 10^{13} \, {\rm GeV}$). The green curve peaks at $1.87 \times 10^{-12}$ and shows the results for only inflationary quantum field fluctuations with low-energy inflation ($H_{\rm inf}^{\rm low} = 2.48 \times 10^9 \, {\rm GeV}$). The red curve peaks at $4.18 \times 10^{-7}$ and represents the results obtained when incorporating adiabatic temperature fluctuations. The analytic approximations for all cases are depicted using dotted lines.

Our analysis highlights a substantial disparity in the influence of adiabatic temperature fluctuations compared to quantum fluctuations. In models characterized by high inflationary scales, where quantum fluctuations are maximized, adiabatic temperature fluctuations are significantly less influential, trailing by approximately three orders of magnitude on large scales. When $f_{\rm a}/H_{\rm inf} \lesssim 1.25 \times 10^4$, the adiabatic temperature fluctuations remain minimal compared to quantum fluctuations. However, as $f_{\rm a}/H_{\rm inf} \gtrsim 1.25 \times 10^4$, adiabatic temperature fluctuations become particularly significant, surpassing quantum fluctuations by up to five orders of magnitude in models characterized by low inflationary scales, and the power spectrum of axion density fluctuations becomes fully dominated by the QCD epoch instead of the physics of inflation. These fluctuations, peaking at a scale of $\sim 6 \times 10^{-7} \, {\rm pc}$ and manifesting after the primordial phase of inflation, exhibit potential for evolving into structures reminiscent of AMC. The typical mass of AMC is predicted to be of the order of $\sim 10^{-13} \, {\rm M}_{\rm \odot}$ with a radius of the order of $\sim 10^{-8} \, {\rm pc}$ in various cosmological models and simulations~\cite{Hogan:1988mp, Kolb:1993zz, Kolb:1994fi, Kolb:1995bu, Xiao:2021nkb}. This transformative evolution is orchestrated by the intricate interplay of gravitational instability, cosmic expansion, and the hierarchical formation of structures. The nuanced dynamics are deeply intertwined with the influence of dark matter, specifically axions, and their gravitational interactions.

\begin{figure}[t!]
\centering
\includegraphics[width=0.65\textwidth]{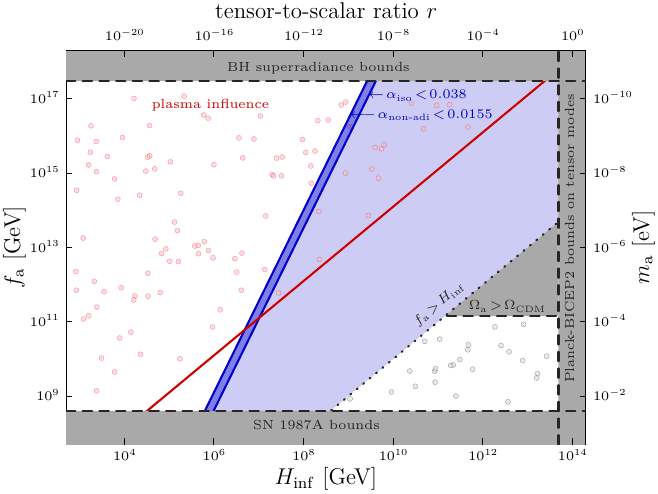}
\caption{
Allowed QCD axion parameter space and dominant mechanisms of seeding axion density fluctuations: 
$H_{\rm inf}$ on the horizontal axis at the bottom, and its corresponding tensor-to-scalar ratio $r$ in single-field slow-roll inflation at the top; PQ breaking scale, $f_{\rm a}$ on the left vertical axis and its corresponding axion mass $m_{\rm a}$ for the pre-inflationary scenario on the right axis (for the post-inflationary scenario this relation does not hold). 
The isocurvature exclusion region, based on Planck Collaboration~\cite{Planck:2018jri} constraints on correlated ($\alpha_{\rm iso}$) and uncorrelated ($\alpha_{\rm non{\text{-}}adi}$) isocurvature perturbations with the adiabatic perturbations, is depicted in light blue. Above the red line the adiabatic temperature fluctuations dominate over the inflationary quantum fluctuations in seeding axion density fluctuations. The black dotted line indicates the domain of the post-inflationary scenario ($f_{\rm a} \lesssim H_{\rm inf}$). Below that line in the post-inflationary scenarios there is a region where axions are overproduced ($\Omega_{\rm a} \gtrsim \Omega_{\rm CDM}$), based on calculations from~\cite{Hannestad:2009ve}. Gray regions, bounded by black dashed lines, indicate the Planck-BICEP2 upper bound on $H_{\rm inf}$ for single-field slow-roll inflation models~\cite{Planck:2018jri}, the upper bound on $f_{\rm a}$ from BH superradiance~\cite{Arvanitaki:2009fg, Arvanitaki:2010sy, Arvanitaki:2014wva}, and the lower bound on $f_{\rm a}$ from SN 1987A observations~\cite{Mayle:1987as}. The grey circles indicate the traditional domain of AMC, the red circles indicate the regime in which thermal fluctuations induce axion density fluctuations peaking at similar comoving scale.}
\label{fig:H_inf_f_a}
\end{figure}

In Figure~\ref{fig:H_inf_f_a}, we illustrate the exclusion and sensitivity regions in the plane of the Hubble scale during inflation, $H_{\rm inf}$, and the PQ breaking scale, $f_{\rm a}$, for the pre-inflationary scenario. The isocurvature exclusion region, based on Planck Collaboration~\cite{Planck:2018jri} constraints on correlated ($\alpha_{\rm iso}$) and uncorrelated ($\alpha_{\rm non{\text{-}}adi}$) isocurvature perturbations with the adiabatic perturbations, is depicted in light blue. Above the red line, adiabatic temperature fluctuations dominate over quantum fluctuations. Thus, in most of the allowed parameter space for the pre-inflationary scenario, the axion field is expected to show a band power spectrum that peaks at the scale associated with the onset of axion field oscillations, and its amplitude is fixed by the adiabatic temperature fluctuations of the QCD plasma in the early Universe. We note that the position and slope of the red line are accurately estimated at the highest values of $f_{\rm a} \sim 10^{16} \, {\rm GeV}$, given the limitations of our step-function approximation of a sudden jump in the axion mass. At lower PQ scales, where the axion mass grows more gradually, this approximation becomes less valid, and the slope of the red line may deviate from this estimate. Black dotted lines indicate the post-inflationary scenario ($f_{\rm a} \lesssim H_{\rm inf}$) and the region where axions overproduce total CDM ($\Omega_{\rm a} \gtrsim \Omega_{\rm CDM}$), based on calculations from~\cite{Hannestad:2009ve}. Gray regions, bounded by black dashed lines, indicate the Planck-BICEP2 upper bound on $H_{\rm inf}$ for single scalar field inflation models, the upper bound on $f_{\rm a}$ from black-hole superradiance~\cite{Arvanitaki:2009fg, Arvanitaki:2010sy, Arvanitaki:2014wva}, and the lower bound on $f_{\rm a}$ from supernova 1987A observations~\cite{Mayle:1987as}.

This analysis elucidates the significant role of adiabatic temperature fluctuations in the pre-inflationary scenario, especially when compared to quantum fluctuations, and underscores the potential of these fluctuations to influence the formation of AMC. Furthermore, our study not only provides valuable insights into the influence of adiabatic and quantum fluctuations on axion density fluctuations but also offers practical advantages in estimating the power spectrum. 
Our approach enables expedited decision-making and data interpretation, making it a highly efficient method for analyzing the power spectrum of axion density fluctuations. 

In our pursuit of a comprehensive understanding of axion density perturbations, we developed a custom numerical code in Python. This code offers a unique blend of simplicity and computational efficiency while maintaining a high level of accuracy. Leveraging insights from publicly available codes~\cite{Mulryne:2016mzv, Visinelli:2009kt}, our code simulates the evolution of the axion field and calculates key quantities such as the axion density, density fluctuations, and power spectrum. The code numerically solves the equations of motion for both the background fields and the evolution of quantum and induced perturbations. 
Additionally, we recognize the challenge posed by the consideration of correlated modes in our analysis. Accounting for the correlation between modes is crucial for a more accurate quantification of the role of adiabatic temperature fluctuations in AMC formation and their impact on axion density perturbations. While our current work assumes uncorrelated modes, we acknowledge that this assumption may not fully capture the complexity of the realistic scenario, especially at a later stage when the CDM perturbations continue to grow logarithmically during the radiation dominated epoch and at small scales start to become non-linear. Therefore, it is important for future research to address this aspect and investigate the effects of mode correlation, thereby providing a more comprehensive understanding of the topic at hand.

Our approach adds depth to our understanding of axion-related phenomena, highlighting the significance of adiabatic temperature fluctuations in shaping the evolution of axion density perturbations. This perspective complements the work in Ref.~\cite{Enander:2017ogx}, which explores scenarios where PQ symmetry breaking occurs after inflation in which the axion field takes on random values in causally disconnected regions, resulting in fluctuations and the formation of AMC. While this work provides valuable insights into the distribution of AMC concerning mass and size, our study offers a unique angle by investigating the impact of adiabatic temperature fluctuations, thus contributing to a more comprehensive understanding of axions as potential dark matter candidates.

\section{Conclusion} \label{sec:7}

In conclusion, axion-like particles (ALPs), including QCD axions, emerge as compelling candidates for explaining the cold dark matter (CDM) content of the Universe. The production of axions through the misalignment mechanism can occur in two scenarios: post-inflationary and pre-inflationary. Discerning the correct scenario provides valuable insights into the scale at which the symmetry-breaking of axions occurred. In the post-inflationary scenario, axions can give rise to high-density configurations during the QCD epoch, evolving into axion miniclusters (AMC) that are gravitationally bound. The identification of these AMC has been proposed as compelling evidence for the post-inflationary scenario.


We showed in this work that adiabatic temperature fluctuations prevail over inflationary quantum fluctuation 
in a large fraction of the allowed parameter space $(H_{\rm infl}, f_{\rm a})$, given by the condition
$f_{\rm a}/H_{\rm inf} \gtrsim 1.25 \times 10^4$. 
These plasma induced axion density perturbations arising after the primordial phase of inflation peak at comoving scales of $\sim 6 \times 10^{-7} \, {\rm pc}$, highlighting the importance of considering these influences, especially in the formation of AMC with typical contemporary scales of $\sim 10^{-8} \, {\rm pc}$ within the pre-inflationary scenario. Our findings challenge the previous notion that the detection of AMC alone can reliably differentiate between the pre-inflationary and post-inflationary origins of axions. Neglecting the influence of adiabatic temperature fluctuations can mask the formation of AMC in the pre-inflationary scenario, particularly in low-energy inflation models. Therefore, the presence of AMC should not be solely relied upon as a discriminator between the two scenarios.

Note that we cannot make any statements about how compact structures seeded during the QCD epoch via coupling to adiabatic density fluctuations would manifest today. The degree of compactness might still differ significantly from the post-inflationary AMC scenario and should be the subject of future studies.

However, it is essential to acknowledge the limitations of our approach. Simplifications and approximations, such as 
neglecting non-linearities 
on small scales are inherent in our analysis. 
The omission of metric perturbations, while simplifying our study, introduces another level of approximation. 
Despite these limitations, our study provides valuable insights into the significance of adiabatic temperature fluctuations in understanding the evolution of axion density perturbations within the context of the pre-inflationary scenario.


In summary, our study underscores the importance of considering adiabatic temperature fluctuations in the study of axions and their cosmological implications, particularly in the pre-inflationary scenario. By demonstrating the significant influence of adiabatic temperature fluctuations on the density perturbations of axions, we have shown that their inclusion is crucial for a comprehensive understanding of the formation of AMC and distinguishing between different production scenarios. Future research, combining theoretical analyses and observational data, will be crucial in further elucidating the role of adiabatic temperature fluctuations and their impact on the CDM puzzle.

\acknowledgments

The authors express their sincere gratitude to Yuko Urakawa, Geoff Beck, and Vladimir Lenok for their invaluable discussions and constructive feedback on an earlier draft of this manuscript. This work is supported by the Deutsche Forschungsgemeinschaft (DFG, German Research Foundation) through the CRC-TR 211 ``Strong-interaction matter under extreme conditions'' (project number 315477589 -- TRR 211).

\normalem
\bibliographystyle{JHEP}
\bibliography{references}
\end{document}